\documentclass[11pt, a4paper]{article}

\usepackage{times}  
\usepackage{helvet}  
\usepackage{courier}  
\usepackage[hyphens]{url}  
\usepackage{graphicx} 
\usepackage[round]{natbib} 
\usepackage{caption} 
\DeclareCaptionStyle{ruled}{labelfont=normalfont,labelsep=colon,strut=off} 
\usepackage{algorithm}

\usepackage{newfloat}
\usepackage{listings}
\floatstyle{ruled}
\newfloat{listing}{tb}{lst}{}
\floatname{listing}{Listing}
\usepackage{graphicx}%
\usepackage{multirow}%
\usepackage{amsmath,amssymb,amsfonts}%
\usepackage{amsthm}%
\usepackage{mathrsfs}%
\usepackage[title]{appendix}%
\usepackage{xcolor}%
\usepackage{textcomp}%
\usepackage{manyfoot}%
\usepackage{booktabs}%
\usepackage{algorithm}%
\usepackage{algorithmicx}%
\usepackage{algpseudocode}%
\usepackage{listings}%

\usepackage{hyperref}

\usepackage[utf8]{inputenc}
\usepackage[font=small]{caption}
\usepackage{subcaption}
\usepackage{amsmath}
\usepackage{braket}
\usepackage{lipsum}
\usepackage[font=small,justification=centering]{caption}
\usepackage{subcaption}
\usepackage{booktabs}
\usepackage{xcolor}
\usepackage{amsfonts}

\usepackage{xr}
\usepackage{fullpage}

\newcommand{\dc}[1]{{\color{black} #1}}

\renewcommand{\cite}[1]{\citep{#1}}

\begin{document}

\title{Peptide Binding Classification on Quantum Computers}

\author{
  Charles London$^{1\dagger}$~~~~
  Douglas Brown$^{1\dagger}$~~~~
  Wenduan Xu$^1$~~~~
  Sezen Vatansever$^2$\\
  Christopher James Langmead$^2$~~~~
  Dimitri Kartsaklis$^1$~~~~
  Stephen Clark$^1$\\
  Konstantinos Meichanetzidis$^1$ \vspace{0.1cm}\\
  \textsf{\{\small charles.london; douglas.brown; wenduan.xu; dimitri.kartsaklis; steve.clark; k.mei\}@quantinuum.com}\\
  \textsf{\{\small svatanse; clangmea\}@amgen.com}\vspace{0.2cm}\\
  $^1$Quantinuum, 17 Beaumont St., Oxford, OX1 2NA, UK\\
  $^2$Amgen, 1 Amgen Center Dr., Thousand Oaks, 91320, CA, USA
}

%

\date{}
\maketitle

\begin{abstract}
We conduct an extensive study on using near-term quantum computers for a task in the domain of computational biology. By constructing quantum models based on parameterised quantum circuits we perform sequence classification on a task relevant to the design of therapeutic proteins, and find competitive performance with classical baselines of similar scale. To study the effect of noise, we run some of the best-performing quantum models with favourable resource requirements on emulators of state-of-the-art noisy quantum processors. We then apply error mitigation methods to improve the signal. We further execute these quantum models on the Quantinuum H1-1 trapped-ion quantum processor and observe very close agreement with noiseless exact simulation. Finally, we perform feature attribution methods and find that the quantum models indeed identify sensible relationships, at least as well as the classical baselines.
This work constitutes the first proof-of-concept application of near-term quantum computing to a task critical to the design of therapeutic proteins, opening the route toward larger-scale applications in this and related fields, in line with the hardware development roadmaps of near-term quantum technologies.
\end{abstract}



\maketitle

\section{Introduction}


In the rapidly growing field of quantum machine learning (QML), the use of parameterised quantum circuits (PQCs) as machine learning models \cite{MarcelloReview} has found a wide range of applications. PQCs provide a number of advantages as tools for QML, the most important of which are relatively easy implementation on quantum hardware, sufficient expressive power to be applied in several tasks \cite{PhysRevResearch.2.033125}, and perhaps, above all, the ability to be trained with a classical machine learning objective function and optimisation loop. In this work, we build and train PQCs to solve a problem in the domain of computational biology.
The task chosen is that of binary classification of peptides, which are short chains of amino acids, according to their binding affinity to a target molecule.

\begingroup
\renewcommand{\thefootnote}{$\dagger$}
\footnotetext{Both authors contributed equally to this work.}
\endgroup

\dc{Peptide binding plays a crucial role in cellular signalling, protein trafficking, immune response, and oncology, and predicting their binding affinity is a long-standing challenge \cite{Das2013-iu, Wang2019-rv}. Due to the importance of the peptide-MHC interaction for adaptive immunity and the large datasets available for training, in the current study we focus on this peptide binding problem.}
We compare the performance of the quantum models\footnote{Specifically, our models fall under the category of \emph{hybrid} quantum models, in the sense that they are composed of PQCs which in turn may be controlled by neural networks (NNs).}
against that of classical baselines, again of sequential structure,
and find that the simple quantum models we define perform as well as the classical neural models.

We consider this a positive outcome for the following reasons. Firstly, quantum models are still in their infancy and face significant technical challenges such as noise and error rates, that limit their practicality in real-world scenarios. Secondly, the inherently different nature of quantum computing from classical computing clearly implies that the strengths of the two paradigms might lie in different problem domains.
Therefore, demonstrating comparable performance on a typical machine learning task offers encouraging insights into the potential of quantum computing in practical applications.

The modest size of the quantum models we investigate in this work allows us to execute them on quantum processors and correct the effect of noise using standardised error mitigation methods. This establishes the first proof-of-concept experiment involving the application of quantum models to a simple computational biology task on currently available quantum hardware.
Finally, we analyse each individual amino acid's contribution to the binding probability, and we observe that our simple small-scale PQC-based models recover this information at least as well as the classical baselines of similar scales.


In summary, the contributions of this paper are as follows: we conduct an extensive study into using quantum ML models on a computational biology task; we detail a methodology that allows the representation of sequence models on quantum hardware; and finally, we provide results from a proof-of-concept experiment on quantum hardware for the potential of quantum models in the field, by achieving results similar to classical baselines.

\section{Background and related work}
\label{sec:related_work}

\dc {

Major histocompatibility complex (MHC) molecules bind short, perfectly cleaved peptides and display them on the cell surface for recognition by T cells, which gives them a central role in regulating the immune response. In view of this, the binding of antigenic peptides to MHC molecules represents an essential step for cellular immunity and understanding the rules of this phenomenon holds valuable potential in human health applications.

MHC comes in two main variants: MHC Class I (MHC-I) and MHC Class II (MHC-II). MHC-I is encoded by three I loci and expressed on the surface of all nucleated cells, whereas MHC-II can only be expressed in professional antigen-presenting cells \cite{Jensen2007-zc}. In this study, we focus on MHC-I molecules (hereafter referred to as MHC).
MHC mainly binds short peptides with a length of 8–10 amino acids that are generated predominantly from intracellular proteins after these have undergone proteasomal degradation \cite{Bouvier1994-mq}. Then, some of these peptide-MHC complexes are presented on the cell surface for recognition by CD8+ T cells, which stimulate cellular immunity \cite{Unanue2006-mx}. Only a small proportion of endogenous peptides can be presented by MHCs because there are thousands of MHC alleles in the human population, each with specificity for binding a distinct set of peptides. This serves as a control mechanism for antigenic variations in the self-peptidome repertoire \cite{Lundegaard2010-yf}. Since the peptide binding is clearly the most selective part of the antigen presentation pathway, predicting the affinity between a peptide and its binding MHC allele has been of particular interest to explain specific immune responses, such as pathogen elimination, transplant rejection, autoimmunity, or death \cite{Neefjes2013-pb}.

In recent years, many computational methods for predicting the binding of MHC to peptides have been proposed. Various types of features have been explored to develop better prediction tools, such as sequences of the peptide and the receptor, structural information of the bound complex, physicochemical properties of the amino acids, evolutionary information, and word embeddings \cite{Hu239236}. Different types of ML algorithms have been applied to MHC–peptide binding prediction and \citet{Luo2015-ts} comprehensively reviewed the existing ML-based methods and discussed their limitations and challenges. These ML approaches include decision trees, hidden Markov models, regression methods, support vector machines, consensus methods, and the more recent deep learning methods using artificial neural networks.
}
In the broader context of bioinformatics and computational biology, transformer networks currently achieve state-of-the-art results, e.g. the recent advances in protein structure prediction led by AlphaFold \cite{Jumper2021HighlyAlphaFold}.

\section{The data, the task, and the methodology}
\label{sec:task}


\begin{figure}[t]
    \centering
        \includegraphics[scale=0.6]{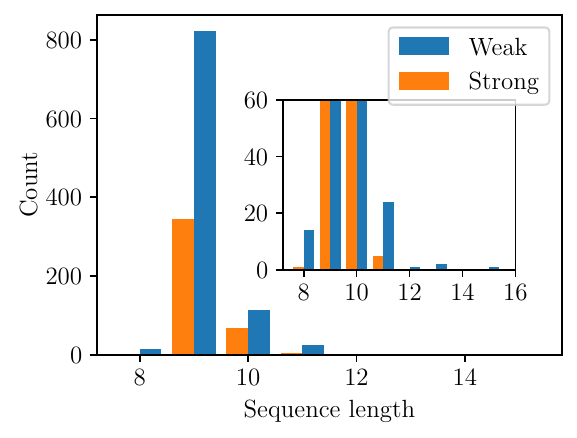}
        \includegraphics[scale=0.6]{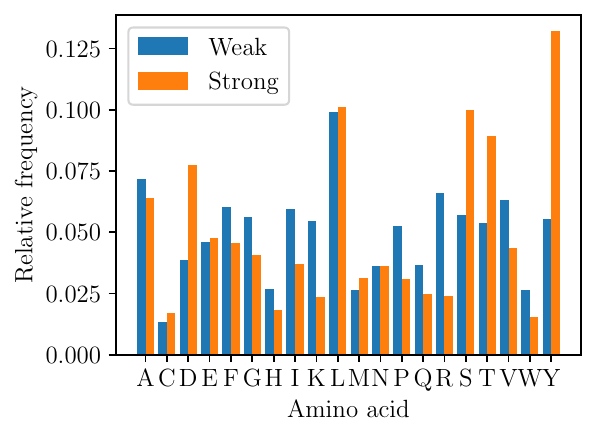}
    \caption{Properties of the peptide data. (Left) Sequence length distribution (inset: zoomed view to show the number of sequences with length $\geq 12$). (Right) Relative distribution of amino acids per class.}
    \label{fig:data_distros_new}
\end{figure}

The MHC Class I binding data were downloaded from the Immune Epitope Database (IEDB) \cite{IEDB} website\footnote{\url{https://www.iedb.org/}}. The entire dataset comprises experimentally measured binding affinities for nearly 200K (peptide, allele) pairs, covering numerous species. For this study, we extracted the 3,237 entries involving the human allele HLA-A*01:01. The choice of allele was arbitrary, but it is found in populations across the globe. Each peptide consists of a sequence of 8-15 amino acids. The amino acids make up a ``vocabulary'' of size $20$. In the original data, each peptide is labelled with its measured half maximal inhibitory concentration ($\mathrm{IC}_{50}$) to HLA-A*01:01. $\mathrm{IC}_{50}$ values are an indirect, but common measure of binding affinity. We transformed the $\mathrm{IC}_{50}$ into $\mathrm{pIC}_{50}=-\log_{10}(\mathrm{IC}_{50})$ values and then applied a threshold of $\mathrm{pIC}_{50}=8$ to create binary labels ($1$ for `strong' and $0$ for `weak'). In summary, the classification task is to predict whether a given 8-15 amino acid peptide binds to HLA-A*01:01.

Since most peptides do \emph{not} bind to HLA-A*01:01, the initial training data are highly imbalanced: 90\% `weak' and 10\% `strong'. To make the data more balanced, we downsample by removing a random subset of the weak peptides to achieve a ratio of 70\% `weak' and 30\% `strong', leaving 1,396 peptides.



Fig. \ref{fig:data_distros_new} shows some statistical properties of the dataset.
The left-hand plot shows the distribution of sequence lengths, and the right-hand plot shows the relative frequency of each amino acid in the two classes. For example, Y makes up 12.5\% of amino acids in strongly bonding peptides, but only 5\% in weakly-bonding ones.
Sequences are overwhelmingly of length $9$, with very few of length $12$ or longer.
The distribution of amino acids is not uniform, nor is the
distribution across the two peptide classes balanced. Some (e.g. $K$, $P$, $R$) appear much more frequently in weakly bonding peptides, and some ($D$, $S$, $T$, $Y$) in strongly binding peptides.

The task at hand comes in the form of a simple supervised binary classification task.
The aim is to train a model such that, given a sequence, it predicts its true binding affinity, `weak' or `strong', with high probability.
We perform $k$-fold cross-validation to get a better characterisation of the generalization ability of the models to unseen data, with $k=5$.
Each fold contains 20\% of the data, and 4 of these folds are used to train the model. The 5th fold is split into a validation set and a test set, giving an 80\%, 10\%, 10\% split, respectively. The validation set is used to select the best weight initialisation and to prevent overfitting in our models via early stopping, and we then
calculate an F1 score for the model on each fold.
More specifically, on each fold, the model is trained with 5 random initialisations and the set of trained weights that performs best on the validation set is chosen. Then, this trained model is applied to the test split of the given fold,
giving a mean and standard deviation of F1 scores over folds,
characterising the performance of the model.



\section{Sequential quantum models} \label{sec:quantum_models}


Our sequential quantum models work by assigning a parameterized quantum circuit with trainable parameters to every amino acid in the vocabulary. By chaining these circuits together in the order of the sequence of amino acids, we obtain a quantum circuit representation of the peptide sequence. Updating the parameters based on a loss allows us to train this circuit in order to classify the binding affinity of each peptide.


\begin{figure}[t]
    \centering
    \includegraphics[scale=0.3]{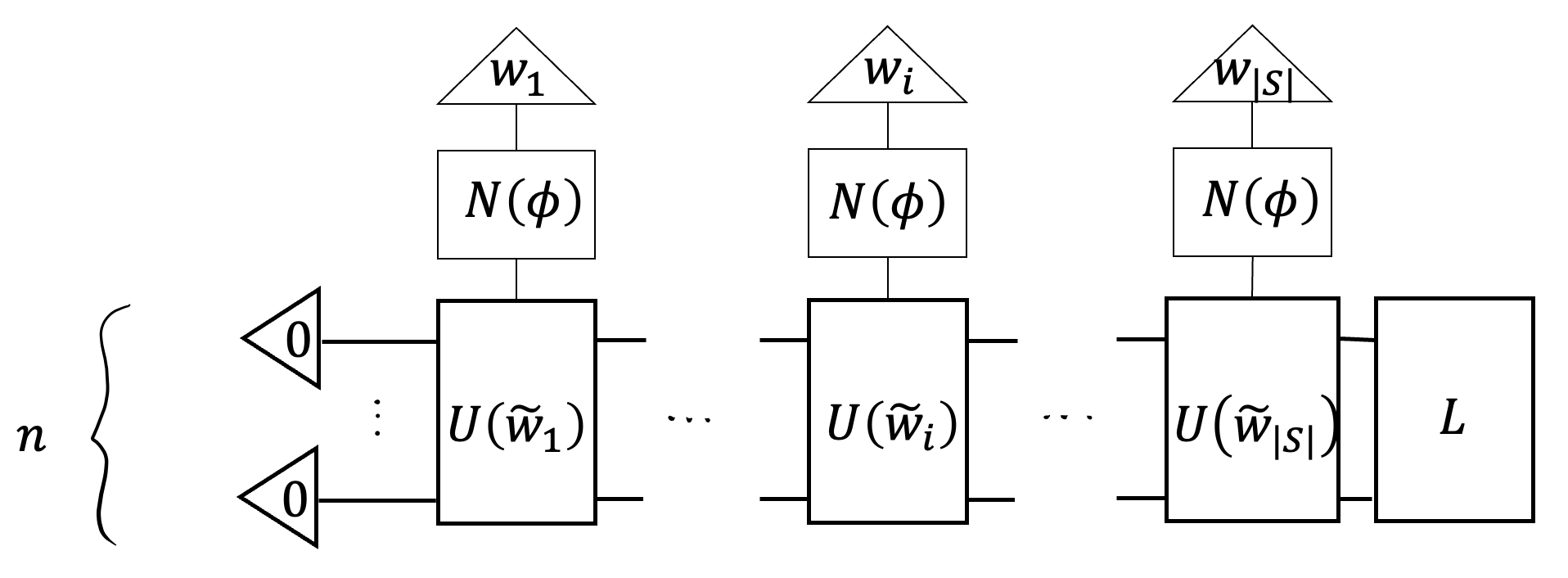}
    \caption{The architecture of the sequential quantum models. Thin wires carry, and thin boxes process, classical information. Thick wires carry, and thick boxes process, quantum information. Amino acid embedding $w_i\in\mathbb{R}^D$ of the $i$-th amino acid in the sequence $S$ is preprocessed by a neural network $N(\phi)$ returning a vector $\tilde{w}_i\in\mathbb{R}^d$, which controls a parameterised quantum circuit $U(\tilde{w}_i)$. The quantum state entering the first amino acid's circuit $U(\tilde{w}_1)$ is $|0\rangle^{\otimes n}$. The $L$-box at the end of the model defines a way to obtain a binary label from the $n$-qubit state exiting $U(\tilde{w}_{|S|})$ (see Fig. \ref{fig:measurements}).}
    \label{fig:seq-q-model}
\end{figure}

\begin{figure}[t]
    \centering
    \includegraphics[scale=0.4]{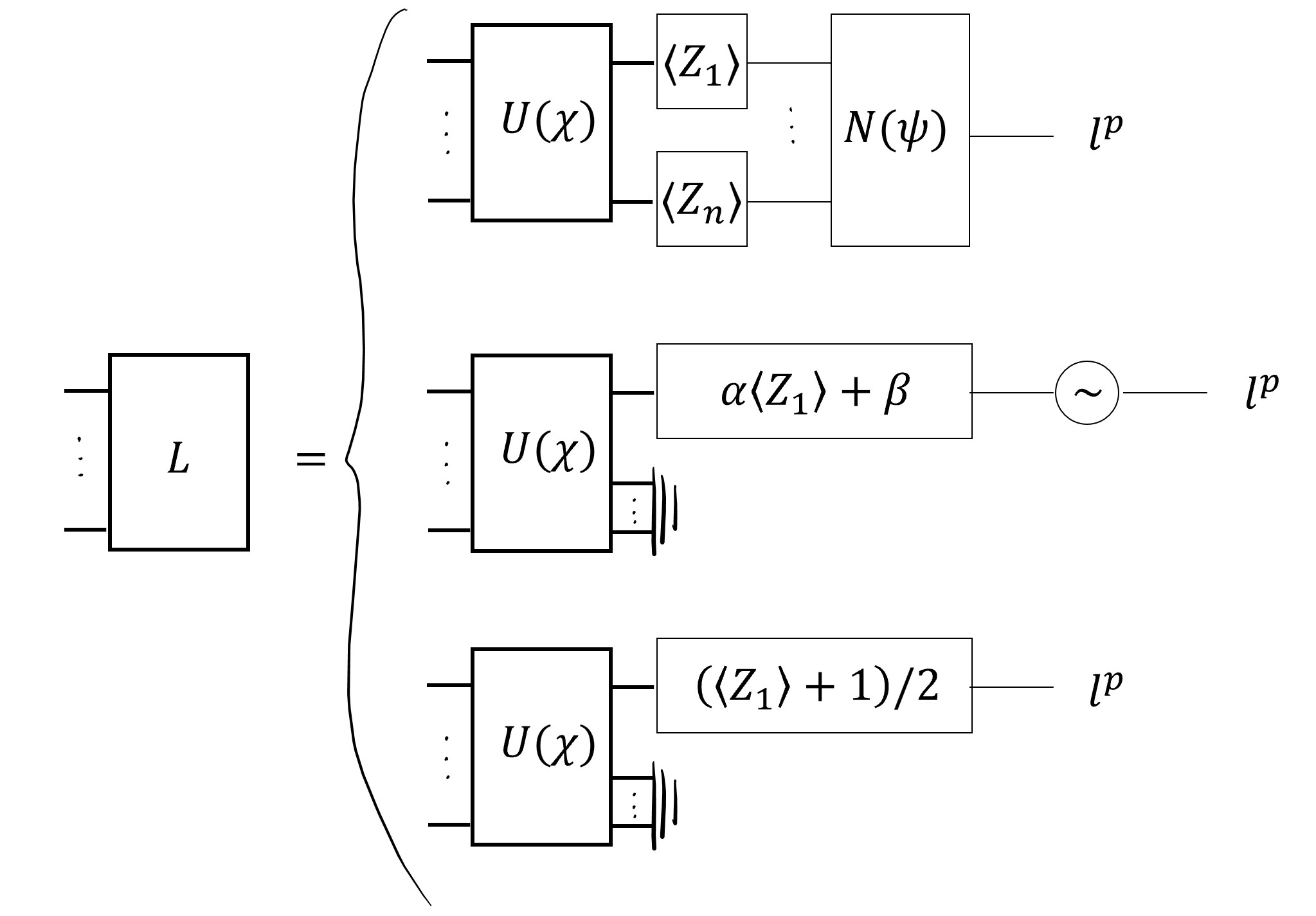}
    \caption{Three ways of obtaining the model's prediction score $l^p\in[0,1]$ from the final state $|\psi\rangle_S$. $U(\chi)$ is a shared classifier among sequences. (Top) $L_1$: measuring $\langle Z_i \rangle$ on all qubits and postprocessing with a neural network. (Middle) $L_2$: Measuring only $\langle Z_1 \rangle$ on the first qubit and multiplying and adding trainable scalars and then applying a sigmoid nonlinearity.
    (Bottom) $L_3$: Measuring only $\langle Z_1 \rangle$ and rescaling so it lies in $[0,1]$.}
    \label{fig:measurements}
\end{figure}

Fig. \ref{fig:seq-q-model} shows the general form of quantum circuits that we will be exploring in this work.
For each peptide sequence $S$ of length $|S|$,
an $n$-qubit quantum state is initialised as $|0\rangle^{\otimes n}\in\mathbb{C}^{2^n}$. Then, $|S|$-many PQCs, $U(\tilde{w}_i)$, one per amino acid, are applied to the state at each time-step, finally returning a state $|\psi\rangle_S$.
Each PQC is, in general, controlled by a parameter set $\tilde{w}_i\in\mathbb{R}^d$, which \emph{depends on the amino acid}.
We have the option to allow neural networks $N(\phi)$, where $\phi$ are the trainable parameters of the network, to learn how to control the PQCs by outputting $\tilde{w}$.These neural networks are given as input an embedding vector $w_i\in\mathbb{R}^D$.
In the case where neural networks are not used, the $\tilde{w}_i$ are simply trained directly using standard backpropagation methods in simulation.
The final state $|\psi\rangle_S$ obtained after the application of all the PQCs on the initial state is then put through a labeling process ($L$-process), which will be defined below, to obtain the model's prediction $l^p \in [0,1]$, which is a score between 0 and 1. If $l^p > 0.5$ then the predicted binary label is 1, otherwise 0.




We explore three choices for the $L$-process from which we obtain the model's prediction.
The first choice uses the Pauli $Z$ expectation values $\langle Z_i \rangle$ from the $n$-qubit state $|\psi\rangle_S$
and
feeds them as an $n$-dimensional vector to a neural network $N(\psi)$,
which outputs a classification probability for label 1.
The Pauli $Z$ expectation on the $i$-th qubit is defined as $\langle Z_i \rangle = P_i(0)-P_i(1) \in [-1,1]$,
i.e. the difference between the probabilities of measuring the $i$-th qubit in the state $0$ or $1$.
The second choice uses only
$\langle Z_1 \rangle$, multiplying this value with a trainable scalar and adding a trainable bias, and finally applying a sigmoid nonlinearity to obtain a label probability.
In the third case, we use only qubit 1 and output the probability of the qubit being in the $\ket{0}$ state (i.e. we measure the observable $\ket{0}\bra{0}$ on qubit 1).
The three choices are illustrated in Fig. \ref{fig:measurements}.

In total, the hyperparameters that specify a trainable model are the following:
the number of qubits $n$, the circuit ansatz used for the PQCs $U$, the number of layers of each ansatz $L_E$, whether the PQC parameters are controlled by a neural network, and the $L$-process.
The choice of which PQCs to use is made based on the expressive ansaetze studied in \citet{Ansaetze}.
In particular, we use Circuits $8$, $9$, and $14$.

Whenever classical neural networks are used in these models, they are fully-connected single-layer perceptrons, with a final sigmoid non-linearity. The size of the neural network is controlled by the dimension $D$ of the embeddings $w_i$ and the dimension $d$ of the embeddings $\tilde{w}_i$. The embeddings $w_i$ can either be pre-trained or trained in-task. We trained the embeddings in task, and randomly initialised them from $\mathcal{N}(0,1)$.

The loss function used for training is the binary cross-entropy loss:
$$H = - \frac{1}{|\mathrm{train ~set}|} \sum_{i=1}^{|\mathrm{train ~set}|} l_i \log_2(l^p_i) + (1-l_i) \log_2(1-l^p_i),$$
where $l_i\in\{0,1\}$ is the correct label for the $i$-th sequence in the train set and $l^p_i\in[0,1]$ is the model's prediction score for that sequence. This formula applies for the complete training set, although in practice we use batches of size 16 during the training process.


In order to evaluate the performance of our models, we choose some appropriate classical baselines to compare against.
The most suitable choices for comparison are sequential models \cite{Goodfellow-et-al-2016} such as recurrent neural networks (RNNs), long short-term memory networks (LSTMs) \cite{HochreiterSchmidhuberLSTM}, and gated recurrent unit networks (GRUs) \cite{cho2014learning}. The parameterisation of these models is given in Table \ref{tab:small_classical} and Table \ref{tab:classical_results_large} in Appendix \ref{secA2}.



\section{Experiments and results}\label{sec:results}

In this section, we present results for a range of quantum models.
The models are defined and trained with exact noiseless simulation, using TorchQuantum (TQ), a PyTorch-based library for hybrid quantum-classical machine learning  \cite{hanruiwang2022quantumnas}.\footnote{The code of the experiments is available at \url{https://github.com/CQCL/peptide-binding-classification-on-quantum-computers/}.}
We also show results from classical baselines with a similar scale (number of trainable parameters). Fairly comparing the expressivity of quantum and classical models is an open area of research, and we use the number of trainable parameters as a coarse approximation.

Further, we select a subset of the quantum models, chosen for their good performance while also exhibiting relatively favourable resource requirements, for execution on quantum computers.
For these, we first obtain \emph{at test time} the F1 scores from emulators of two near-term quantum processors. We then execute one of the models for a subset of the data on an actual quantum computer.

Finally, we apply feature attribution methods in order to identify the relevant amino acids that are responsible for the peptides' classification labels.
We find that the quantum models identify the relevant features in agreement with the ground truth, as clearly as the classical baselines of similar scale.

\dc{Since the peptide binding affinity towards single MHC proteins is a complex function of its amino acid sequence, understanding how much each amino acid in the model contributed to the binding strength predictions for the given MHC can allow the investigation of new aspects of peptide-MHC binding.}

\subsection{Exact classical simulation}\label{sec:exact-classical-simulation}

\begin{table}[t]
\centering
\small
\begin{tabular}{lrlrrrlrll}
\toprule
{} &  NN & Ans &  $L_E$ &  $n$ &    $L$ &  Nº par & Test F1 & SD \\
\midrule
Q1  &      T &      8 &      1 &         4 &  $L_1$ &        841 &    0.80 &    0.04 \\
Q2  &      T &      9 &      1 &         8 &  $L_1$ &        665 &    0.79 &    0.05 \\
Q3  &      T &      8 &      1 &         8 &  $L_1$ &       2505 &    0.79 &    0.04 \\
Q4  &      T &      9 &      1 &         2 &  $L_2$ &        266 &    0.78 &    0.04 \\
Q5  &      T &      9 &      1 &         2 &  $L_1$ &        267 &    0.78 &    0.04 \\
Q6  &      T &      9 &      1 &         4 &  $L_2$ &       1178 &    0.77 &    0.04 \\
Q7  &      F &     14 &     1 &         2 &  $L_1$ &        195 &    0.75 &    0.04 \\
Q8  &      F &     14 &     1 &         2 &  $L_2$ &        194 &    0.75 &    0.04 \\
Q9  &      T &      9 &      1 &         2 &  $L_3$ &        264 &    0.75 &    0.04 \\
Q10 &      T &      9 &      1 &         4 &  $L_1$ &        269 &    0.72 &    0.05 \\
\bottomrule
\end{tabular}

\caption{Results from selected quantum models. For each model, we show by NN whether the PQC parameters are controlled by a neural network, the quantum layers of the ansaetze $L_E$ corresponding to each amino acid, the number of qubits $n$, the method $L_1$, $L_2$, or $L_3$ for obtaining the binary label, and the number of parameters and the F1 score averaged over folds along with its standard deviation.}
\label{tab:quantum_results_large}
\end{table}

\begin{table}[t]
\centering
\small
\begin{tabular}{llrrrrrr}
\toprule
{} & Type & Layers &  Inp. &  Hid. &  Nº par & Test F1 & SD\\
\midrule
K1 &   RNN &          1 &          10 &           20 &        881 &    0.82 & 0.04\\
K2 &   GRU &          1 &          10 &           10 &        891 &    0.79 & 0.03\\
K3 &  LSTM &          1 &           9 &            9 &        928 &    0.79 & 0.04\\
K4 &  LSTM &          1 &           3 &            4 &        215 &    0.77 & 0.05\\
K5 &   RNN &          1 &           5 &            6 &        195 &    0.77 & 0.05\\
K6 &   GRU &          1 &           4 &            4 &        213 &    0.76 & 0.05\\
K0 &   RNN &          1 &           1 &            1 &         28 &    0.76 & 0.06\\
\bottomrule
\end{tabular}
\caption{Results from classical models K$i$ of similar scale with the best performing small-scale quantum model Q1. For each model, we show the model type, the number of layers, the dimensions of the input and hidden layer, the number of parameters and the average F1 score over folds along with the standard deviation.}
\label{tab:small_classical}
\end{table}

\begin{figure}[t]
    \centering
    \includegraphics[scale=0.8]{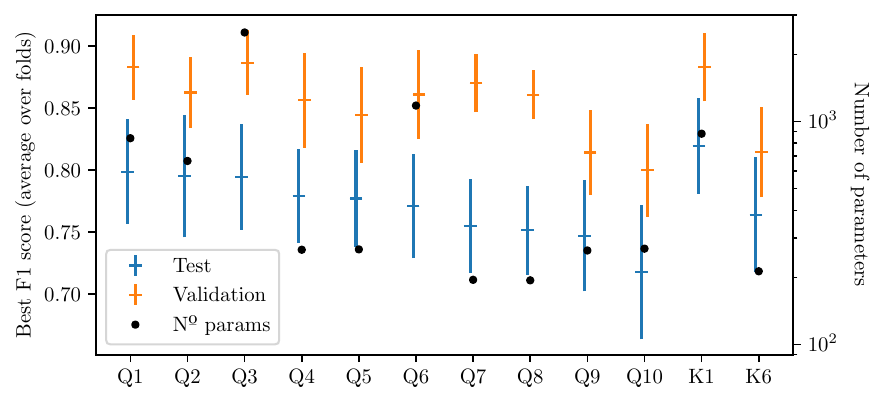}
    \caption{Average F1 score over folds on test and validation data for a range of models. Error bars are standard deviations. All quantum models (Q$i$) tested are shown, as well as the best and worst performing classical models K1 and K6 that are of similar scale to Q1.}
    \label{fig:accuracies_quantum}
\end{figure}

Table~\ref{tab:quantum_results_large} shows the results from a selection of small-scale quantum models.
For each model, the F1 score shown is the average of the test F1 scores for each fold, achieved using noiseless classical simulation with TQ.
The best quantum model tested, Q1, has an average test F1 score across folds of 0.80, with a standard deviation of 0.04. Parameter count does not correlate monotonically with performance. For example, Q1 and Q5 perform better than Q6 despite having many fewer parameters.
We note that increasing $n>8$ did not increase performance, as we observed when training on a subset of the dataset (to reduce training cost) comprising 390 peptides with 40\% `strong' and 60\% `weak'.

Table \ref{tab:small_classical} shows results for the baseline classical models, defined to be of similar scale, i.e. at least the same order of magnitude, to the quantum models whose results are shown in Table~\ref{tab:quantum_results_large}.
By `scale' we refer both to the dimensions of the quantum states and the vectors involved in the quantum and classical models respectively, as well as the total number of trainable parameters.
Notably, the simple quantum models Q$i$ we have defined achieve competitive performance compared to sequential neural models K$i$ with highly established architectures.
In Fig. \ref{fig:accuracies_quantum} we show the F1 scores for all quantum models and the best and worst performing classical baselines of similar scale, together with their number of parameters.
We do not observe any strong correlation between the number of model parameters and the performance at test time.

In Appendix \ref{secA1} we show the corresponding results from larger-scale classical baseline models,
in order to explore the best possible performance achievable by sequential neural models of the same type of architecture.

\subsection{Execution on quantum emulators and devices}

\begin{table}[t]
\centering
\small

\begin{tabular}{llll|ll}
\toprule
{} & \multicolumn{3}{c|}{All folds} & \multicolumn{2}{c}{fold\_0 only} \\
{} & TQ & H1-1E & Aer ibm\_lagos & TQ & H1-1        \\
\midrule
Q1  & 0.80 & 0.80 & 0.78 & 0.76 & 0.75 \\
Q4  & 0.78 & 0.77 & 0.76 & - & - \\
Q10 & 0.72 & 0.72 & 0.68 & - & - \\
\bottomrule
\end{tabular}

\caption{(Left) F1 scores averaged over all folds for selected quantum models on the noiseless simulator TQ and on two different quantum emulators, Quantinuum's H1-1E emulator and IBM's Aer emulator with a noise model from ibm\_lagos. (Right) F1 scores for fold\_0 for TQ and Quantinuum's quantum computer H1-1. In all cases, the number of shots for the emulators and the real device was $2^{10}$.}
\label{tab:emulator_results}
\end{table}


\begin{table}[t]
    \centering
    \small
    \begin{tabular}{lrrr}
\hline
 {}   &   Pre-compiled &   H1-1 &   ibm\_lagos \\
\hline
 Q1        &                 30 &          60 &                  60 \\
 Q4        &                 20 &           3 &                   3 \\
 Q10       &                 30 &          17 &                  35 \\
\hline
\end{tabular}
    \caption{Number of two-qubit gates pre- and post-compilation for the selected quantum models on two backends, Quantinuum's H1-1 and IBM's ibm\_lagos.}
    \label{tab:two_qb_wo_oqc}
\end{table}

Table~\ref{tab:emulator_results} shows the results of selected quantum models run \emph{at test time} on the Quantinuum H1-1E emulator, which runs an accurate, shot-based noise model of the H1-1 quantum processor, and IBM's Aer emulator with a noise model from an IBM quantum device which we choose to be ibm\_lagos.
In all cases, the number of shots used is $2^{10}$.
We also execute the test set of the first fold (fold\_0) of the dataset on the real H1-1 quantum device to demonstrate the performance on actual quantum hardware. Here we find good agreement with the results obtained by the H1-1E emulator.
We observe that all achieve comparable performance to the exact noiseless simulation with TQ.

Upon submitting a circuit for execution on a quantum processor, or its emulator, the circuit first needs to be \emph{compiled}. This involves translating the gates in the abstract circuit being submitted into the native gate set available on the machine. Further, some backends, such as those based on superconducting integrated chips, like ibm\_lagos considered here, have topology constraints, and satisfying these constraints can increase the circuit depth \cite{QubitRouting}.
An illustration of the number of two-qubit gates, which are the lowest fidelity gates on both backends, are shown pre- and post-compilation in Table \ref{tab:two_qb_wo_oqc} for one particular circuit in the test set.
The all-to-all connectivity of H1-1 is favourable for keeping the circuit depth shallow and the number of entangling gates to a minimum. The size of the compiled circuit is affected by the native gateset of the device in question, as well as the properties of the gates in each circuit architecture and how they combine when stacked in series. 

\begin{figure}[t]
    \centering
    \includegraphics[scale=0.8]{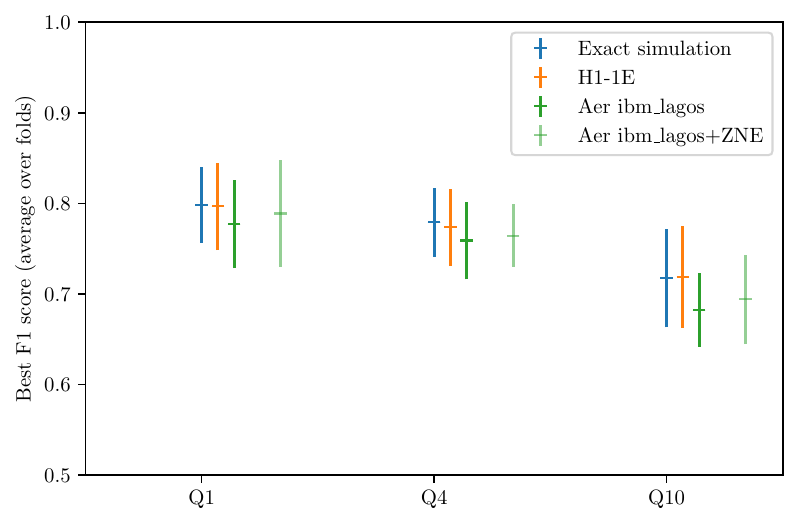}
    \caption{F1 score and standard deviation across folds at test time for the three selected quantum models on the emulators H1-1E and Aer with ibm\_lagos' noise model. Error mitigation with the ZNE method improves the result obtained from Aer with ibm\_lagos' noise model to match that of the ideal exact noiseless simulation and the emulation on H1-1E.}
    \label{fig:emulator_f1s_with_oqc_with_qermit}
\end{figure}

To showcase the feasibility of executing our quantum models on noisy backends, we have also used the error-mitigation package Qermit \cite{cirstoiu2022volumetric} to improve the signal obtained by the emulation of ibm\_lagos.
Error mitigation refers to various techniques which can be used on currently available quantum processors to reduce the noise in the expectation values produced. The method we apply here is known as zero-noise extrapolation (ZNE), whose overhead is only linear in the size of the circuit being executed. Noise in a circuit is artificially increased, and the resulting expectation values are extrapolated backwards to estimate the zero-noise case. Specifically, the noise is increased by replacing a CNOT gate with the sequential application of an odd number of CNOTs, since an even number of CNOTs cancel out. Here, the noise is scaled by factors of 3, 5 and 7, and the ZNE method is found to improve the result obtained by the emulation of ibm\_lagos, recovering the result obtained by classical simulation with TQ.
Fig.~\ref{fig:emulator_f1s_with_oqc_with_qermit} shows the F1 scores obtained at test time for the three selected quantum models, obtained via exact simulation with TQ, emulation on H1-1E, and emulation on Aer with ibm\_lagos' noise model, both with and without ZNE.

\subsection{Determining feature importance}

When performing machine learning classification tasks, we often wish to determine the relative contribution of different input features to the predicted class. This is referred to as \textit{feature attribution} (FA).
In our case, we would like to know how much influence an amino acid at a certain position in the sequence has on the likelihood of the peptide binding to the substrate. This can give us an insight into which structures and chemical groups make a given peptide a more effective binder, suggesting promising directions for designing new molecules with the desired binding properties.

A number of methods have been proposed for FA, but they can generally be split into gradient-based and non-gradient-based methods.
We use the Pytorch package Captum \cite{kokhlikyan2020captum} and apply the gradient-based `integrated gradients' method and the non-gradient-based `Shapley value sampling' method.

\subsubsection{Integrated gradients}

The integrated gradients (IG) method is based on the idea that the contribution of each input feature to the output label can be computed from the gradients of the output with respect to the input.
The method works by first defining a baseline input, usually all zeros, with the same dimensions as the actual input. The baseline input is then gradually changed to the actual input by computing the gradients of the output with respect to the input at each step and integrating them along the path from the baseline input to the actual input. The integrated gradients for each input feature are computed by averaging the gradients computed at each step along the path.

\begin{figure}[t]
\centering
    \includegraphics[scale=0.4]{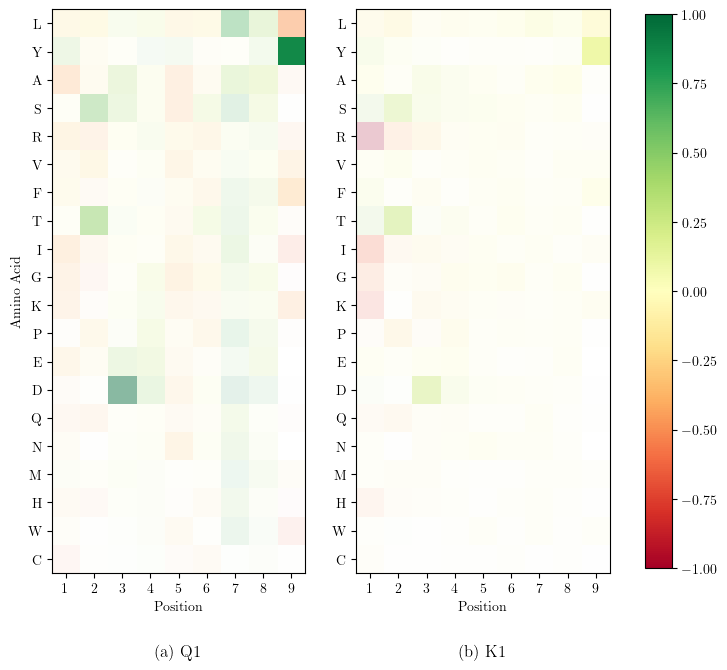}

    \caption{Average IG attributions for each amino acid at each position in peptide sequences of length 9. Positive/negative attribution (green/red) indicates strong/weak binding affinity. Colour saturation indicates the frequency of an amino acid in that position. The features identified by the best-performing quantum model Q1 include the features identified by the best-performing classical model K1 of similar scale.}

\label{fig:ig_att}
\end{figure}

The resulting integrated gradients represent the contribution of each input feature to the output. Highly positive integrated gradients indicate that the input feature increases the output value, while highly negative integrated gradients indicate that the input feature decreases the output value. The integrated gradients can be visualized as a heatmap over the input features, providing a clear and intuitive understanding of which features are most influential in the network's output.
Fig. \ref{fig:ig_att} displays the attribution values computed by integrated gradients for the best quantum and classical models, averaged over all occurrences of a given amino acid in a given position. We restrict this analysis to sequences of length 9, as these make up the overwhelming majority of our dataset (see Fig. \ref{fig:data_distros_new}), and this removes the additional complication of factoring in the distance of each amino acid from both the beginning and the end of the sequence. The colour of each square shows the direction and strength of each attribution, with dark green signifying a significantly positive attribution (corresponding to a `strong' binding affinity), and dark red a significantly negative attribution (corresponding to a `weak' binding affinity), which we normalise to the range $[-1, 1]$ by dividing by the maximum absolute value of the average attributions. The saturation of each square displays the frequency that an amino acid occurs in that position, normalised to the range $[0, 1]$ by dividing by the count of the most frequent combination (Y, 8 in this case).

From Fig. \ref{fig:ig_att} we can see that the two models Q1 and K1 of similar scale largely agree on the three most important features, with (Y, 8), (D, 2), and (T, 1) all having a positive impact on the binding probability of a peptide, although Q1 is more confident about the importance.
There is also agreement on the negative effect of (K, 0), (G, 0), (I, 0), and (R, 0), although K1 is more confident.
There are some relatively strong attributions discovered by the quantum model that seem to be missed by the classical.
Chief among these are (L, 6) and (L, 8), which Q1 determines to have significantly stronger attributions than K1 does.

\subsubsection{Shapley value sampling}

Shapley value is a concept from cooperative game theory that can be used to compute feature attributions for machine learning models. If we consider the output of a machine learning model as a function of its input features, then the Shapley value of each input feature is the average marginal contribution of that feature to the output across all possible orderings of the input features. To compute the exact Shapley values, we would need to compute the output of the model for every possible ordering of the input features, by adding each feature one by one to a baseline input (in this case the all-zero input). The marginal contribution of a feature to each ordering is the difference in output before and after that feature is added. By averaging the contribution of each feature across all possible orderings, we obtain the Shapley value of that feature. However, the number of possible orderings grows exponentially with the number of input features, which makes this intractable to calculate for all but the smallest models. Shapley value sampling computes a Monte Carlo estimate of the Shapley values by randomly sampling $m$ permutations of features, computing the marginal contribution of each feature to each permutation, and averaging the contributions across all sampled permutations. The number of permutations $m$ is a hyperparameter that can be tuned to trade off accuracy for computational cost, and we use $m=25$. Non-gradient-based attribution methods, such as SVS (Shapley Value Sampling) and feature ablation, have the significant advantage of still being accurate even when exact gradients are impossible to obtain, as would be the case when running large instances of quantum models on actual quantum hardware.

\begin{figure}[t]
\centering

    \includegraphics[scale=0.4]{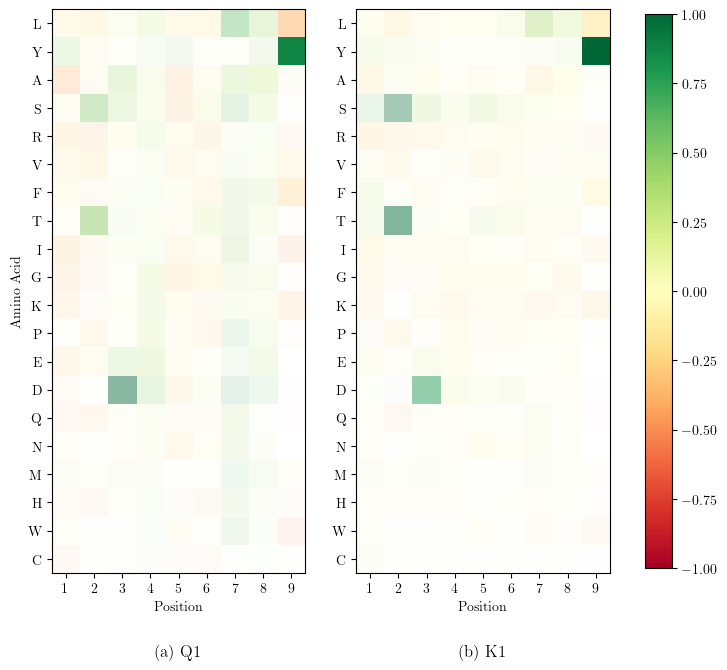}

    \caption{Average SVS attributions for each amino acid at each position in peptide sequences of length 9. Positive/negative attribution (green/red) indicates strong/weak binding affinity. Colour saturation indicates the frequency of an amino acid in that position. The features identified by the best-performing quantum model Q1 include the features identified by the best-performing classical model K1 of similar scale.}

\label{fig:svs_att}
\end{figure}

The resulting Shapley values represent the contribution of each input feature to the output. Highly positive Shapley values indicate that the input feature increases the output value, while highly negative Shapley values indicate that the input feature decreases the output value.
In Fig. \ref{fig:svs_att} we display the average attribution results when using SVS to compute attributions,
following the same plotting conventions as in Fig. \ref{fig:ig_att}.
SVS seems to provide a clearer indication of the dominant features when applied to K1.
Q1 again seems to pick up some features ignored by K1, namely (L, 8) and (A, 0).

The agreement between the SVS and IG attributions regarding the most relevant features, as recovered from both models Q1 and K1, should inspire confidence in the correctness of the attributions discovered.
To increase our confidence about at least the dominant features,
in Appendix \ref{secA2},
we also apply both FA methods to the best performing of the larger-scale classical models, namely C1.
We find that the dominant features found by both FA methods on C1 are indeed in agreement with those found by both FA methods on Q1 and by SVS on K1.
Interestingly, the IG method provides less clear results on K1 than it does on Q1 or C1, even if the quantum model Q1 is of a much smaller scale than C1.

\dc{The binding affinity between a peptide and its specific MHC molecule is largely determined by key amino acids: anchor residues at both peptide termini, usually positions 2 (P2) and 9 (P9) of the 9-mer ligands \cite{Saito1993-ec, Antunes2018-hp}. These residues have been named ``anchors'' due to their ability to fit in ``pocket'' within the groove of the MHC molecule \cite{Hobohm1993-an}. Allele-specific binding pockets favor certain anchor residues (e.g., P2 and P9) and provide peptide ligand specificity for polymorphic MHC-I molecules \cite{Zajonc2020-ni}. Since changes in primary peptide anchor residues can substantially alter MHC binding, improving peptide antigens by altering MHC anchor residues is a common strategy used to enhance binding affinity \cite{Hamley2022-md}. Therefore, it is important to know the anchor's amino acid residue preference.

Our feature extraction analysis discovered strong attributions at the anchor positions. Specifically, our models identified the preference of an aromatic amino acid (Y) at position 9, which conforms to the biological relevance affirmed by previous studies that aromatic residues tend to be favored in the C-terminal position \cite{Zajonc2020-ni, Wieczorek2017-zp}. The enrichment of Y at the ninth position was also shown in a peptide presentation dataset \cite{Lawrence2022-qt}. Also noteworthy is that Q1 and K1 models revealed the enrichment of the central amino acids (position 3) indicating that models were able to learn some motifs that convey binding favorability within the central amino acids. Furthermore, the high representation of D at position 3 is consistent with the previous findings.
}

\section{Discussion and future work}\label{sec:conclusion}

We have constructed hybrid quantum models with a focus on biosequence classification,
and have performed the first proof-of-concept experiments for such applications on a trapped-ion quantum processor. We have found that execution on real quantum hardware shows good agreement with results obtained from exact classical simulation and emulation of the device with an accurate noise model.
Finally, feature extraction methods successfully identified the relevant amino acids responsible for the binding affinity of the peptides being classified.
\dc{MHC variants influence many important biological traits, and the varied peptide binding specificity of these highly polymorphic molecules has important consequences for vaccine design, transplantation, autoimmunity, and cancer development. However, the specific role of peptide-repertoire variation in each variant is not well understood since the characterization of the peptide binding preferences of individual MHC proteins is experimentally challenging. To fill in this knowledge gap, our approach integrates the prediction of peptide binding strength and the identification of residues that influence target-specific binding.}


Concerning the architecture of our models, we note that the circuits involved in this work are one-dimensional (Fig. \ref{fig:seq-q-model}). This means that the computational cost of simulating these quantum models classically would scale exponentially only with the number of qubits $n$, but not with the length of the input sequence. The number of qubits required for reasonable performance for the particular task on the particular dataset in this work was small enough to allow for their efficient simulation. This also allowed for the execution of the models on readily available quantum hardware, where one needs to be conservative with the available quantum resources as quantum technologies are still in their early stages of development.
The work naturally extends to longer sequences, multiclass classification, and classification of tuples of sequences. More interestingly, it is worth exploring tasks for which the problem-native compositional structure would result in circuits with more complex connectivity, similar to work already performed in quantum natural language processing \cite{GrammarAware-KMei,QNLPinPractice}.

\section*{Acknowledgements}

We would like to thank Marcello Benedetti for providing helpful feedback.

\begin{appendices}

\section{Effect of shot noise}

For the main experiment, an average was taken over the cross-validation folds to understand the variability of the performance of the model.
Due to usage demand for the real quantum device, only the test set of the first fold of the data was run on H1-1.
However, we can understand the reliability of this result by considering the shot noise on the expectation values.

The additive error on each expectation value taken at the end of the quantum circuit will be $\sim\frac{1}{\sqrt{N_s}}$, where $N_s$ is the number of shots. The model which was executed on H1-1, Q1, has the labels computed by taking the expectation values of all four qubits and then using a linear layer to get a final output for classification, as described in Section \ref{sec:quantum_models}. The linear layer consists of a single matrix of dimension (4, 1) and a bias.

The effect of the shot noise can be estimated by considering a vector $(\pm \delta, \pm \delta, \pm \delta, \pm \delta)$ where $\delta$ is $\frac{1}{\sqrt{N_s}}$.
An estimate for the effect of noise can be made by adding each of the 16 possible vectors to the output expectations of all circuits in the batch.
The global minimum and maximum F1 scores across these 16 noise-altered cases can then be obtained, illustrating the best- and worst-case results under shot noise.

The original F1 score for this configuration (i.e on fold\_0 with model Q1) with the noiseless TQ simulation was 0.763. The F1 score from H1-1 with unaltered expectation values was 0.747.
Performing the analysis above gives bounds of (0.730, 0.785).

\section{Larger classical baselines}\label{secA1}

Here, we present results from classical neural baseline models, C$i$, of the same architecture as the smaller baselines K$i$ presented in the main text. The C$i$ models are not restricted to be of similar scale to the quantum models Q$i$.

\begin{figure}[t]
    \centering
    \includegraphics[scale=0.8]{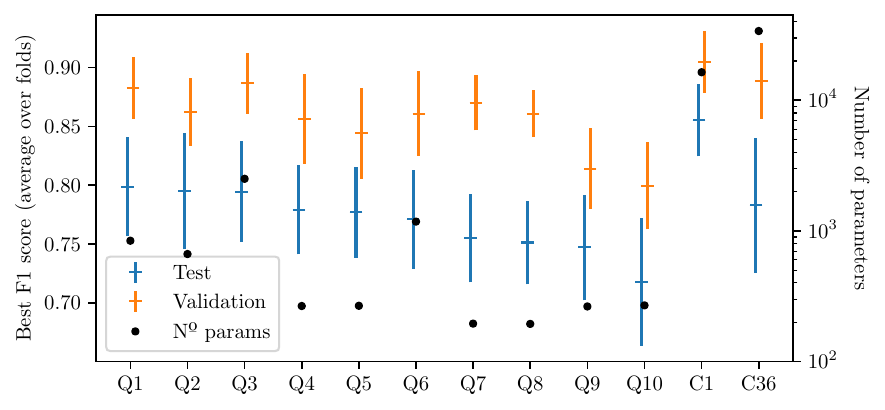}
    \caption{Average F1 score over folds on test and validation data for a range of models. All quantum models (Q$i$) tested are shown, as well as the highest- and lowest-scoring larger-scale classical models (C$i$).}
    \label{fig:accuracies_quantum-C}
\end{figure}

The methodology followed to obtain these results is identical to that followed for the results shown in the main text.
In Table \ref{tab:classical_results_large} we show the average F1 score over folds for these larger models C$i$,
and in Fig. \ref{fig:accuracies_quantum-C} we show these F1 scores for the best and worst performing large-scale classical baselines, C1 and C36, along with the small scale quantum models Q$i$ presented in Section \ref{sec:exact-classical-simulation} the main text.

\begin{table}[t]
\small
\centering
\begin{tabular}{llrrrrl}
\toprule
{} & Model &  layers &  Inp. &  Hid. &  Nº par & Test F1 \\
\midrule
C1  &   RNN &          1 &          50 &          100 &      16401 &    0.86 \\
C2  &   RNN &          2 &          50 &           50 &      11351 &    0.85 \\
C3  &   RNN &          1 &          20 &          100 &      12741 &    0.85 \\
C4  &   RNN &          2 &          20 &          100 &      32941 &    0.85 \\
C5  &   RNN &          2 &          20 &           50 &       9191 &    0.85 \\
C6  &   RNN &          2 &          50 &          100 &      36601 &    0.85 \\
C7  &   RNN &          2 &          10 &           50 &       8471 &    0.84 \\
C8  &  LSTM &          2 &          50 &          100 &     142801 &    0.83 \\
C9  &   GRU &          1 &          50 &          100 &      46801 &    0.83 \\
C10 &   RNN &          1 &          10 &           50 &       3371 &    0.83 \\
C11 &   RNN &          1 &          20 &           50 &       4091 &    0.83 \\
C12 &   GRU &          2 &          50 &          100 &     107401 &    0.83 \\
C13 &   RNN &          2 &          10 &          100 &      31721 &    0.83 \\
C14 &  LSTM &          1 &          50 &          100 &      62001 &    0.83 \\
C15 &   GRU &          2 &          10 &          100 &      94521 &    0.82 \\
C16 &   RNN &          1 &          10 &          100 &      11521 &    0.82 \\
C17 &  LSTM &          2 &          20 &           50 &      35291 &    0.82 \\
C18 &   GRU &          2 &          20 &          100 &      97741 &    0.82 \\
C19 &   GRU &          2 &          20 &           50 &      26591 &    0.82 \\
C20 &   GRU &          1 &          20 &          100 &      37141 &    0.82 \\
C21 &  LSTM &          2 &          50 &           50 &      41951 &    0.82 \\
C22 &   RNN &          1 &          50 &           50 &       6251 &    0.82 \\
C23 &   GRU &          2 &          50 &           50 &      31751 &    0.81 \\
C24 &   GRU &          1 &          20 &           50 &      11291 &    0.81 \\
C25 &  LSTM &          1 &          20 &          100 &      49341 &    0.81 \\
C26 &  LSTM &          2 &          10 &           50 &      33071 &    0.81 \\
C27 &   GRU &          1 &          10 &           50 &       9571 &    0.81 \\
C28 &  LSTM &          1 &          20 &           50 &      14891 &    0.81 \\
C29 &   GRU &          2 &          10 &           50 &      24871 &    0.81 \\
C30 &   GRU &          1 &          50 &           50 &      16451 &    0.81 \\
C31 &  LSTM &          2 &          10 &          100 &     125921 &    0.81 \\
C32 &  LSTM &          2 &          20 &          100 &     130141 &    0.80 \\
C33 &  LSTM &          1 &          50 &           50 &      21551 &    0.80 \\
C34 &  LSTM &          1 &          10 &          100 &      45121 &    0.80 \\
C35 &  LSTM &          1 &          10 &           50 &      12671 &    0.79 \\
C36 &   GRU &          1 &          10 &          100 &      33921 &    0.78 \\
\bottomrule
\end{tabular}

\caption{Results from classical models C$i$. For each model we show the model type, the number of layers, the dimensions of the input and hidden layer, the number of parameters and the best F1 score averaged over folds.}
\label{tab:classical_results_large}
\end{table}

\section{Feature attribution for larger classical baselines}
\label{secA2}

\begin{figure}[t]
\centering
    \includegraphics[scale=0.4]{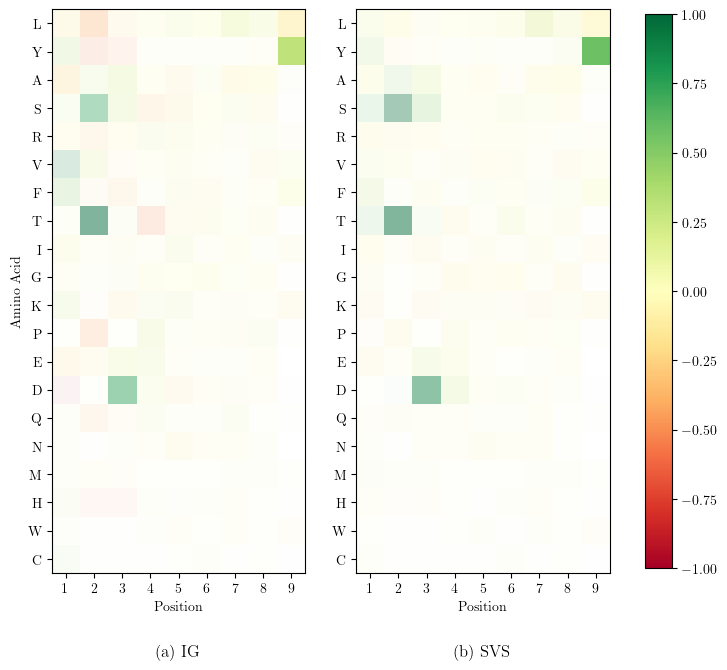}

    \caption{Average IG and SVS attributions for each amino acid at each position in peptide sequences of length 9. Positive/negative attribution (green/red) indicates strong/weak binding affinity. Colour saturation indicates frequency of an amino acid in that position. For the best-performing classical larger-scale model C1.}

\label{fig:ig_svs_att-larger}
\end{figure}

Here we show heatmaps obtained by the two FA methods introduced in the main text, IG and SVS, for the best-performing large-scale classical model C1 (see Appendix \ref{secA1}).

From Fig. \ref{fig:ig_svs_att-larger} we can see that model C1 identifies as most important features (Y, 8), (D, 2), and (T, 1), as all having a large positive impact on the binding probability of a peptide. IG also identifies that (S, 1) and (A, 0) have a positive and negative impact, respectively, albeit with low importance.
SVS identifies (S, 1) as more significant than IG does, and also identifies (L, 6) as positive but quite low importance.

Interestingly,
the larger model C1 identifies more clearly the relevant features than the smaller scale model K1, especially when the IG method of FA is used.
However, for the IG method, the quantum model Q1 identifies the relevant features more clearly than K1 does, even if smaller scale than C1 (see Figs. \ref{fig:ig_att} and \ref{fig:svs_att} in the main text).

\end{appendices}

\bibliographystyle{plainnat}
\bibliography{refs}

\end{document}